\documentclass[useAMS,usenatbib]{mn2e}
\usepackage{natbib, aas_macros}
\usepackage{version}
\usepackage{ulem}
\usepackage{graphicx}
\usepackage{wrapfig}
\graphicspath{{./figs/}}
\usepackage{color}
\usepackage{amsmath}
\usepackage{amssymb}

\newcommand{\Msun}{{\rm  M_{\odot}}}

\newcommand{\Zsun}{Z_{\odot}}

\newcommand{\lya}{\rm {Ly{\alpha}}}

\newcommand{\Tvir}{T_{\rm vir}}

\newcommand{\cc}{{\rm cm^{-3}}}

\newcommand{\un}[2]{#1_{\rm #2}}

\newcommand{\HM}{\rm{H_{2}}}

\title[Gas clump formation in high-$z$ dwarf galaxies] %sub-title
{
Gas clump formation via thermal instability in high-redshift dwarf galaxy mergers\\
}
\author[Arata et al.]
{Shohei Arata$^{1}$\thanks{E-mail: arata@vega.ess.sci.osaka-u.ac.jp},
Hidenobu Yajima$^{2,3}$,
and Kentaro Nagamine$^{1,4}$
\\
$^{1}$ 
Department of Earth and Space Science, Osaka University, Toyonaka, Osaka 560-0043, Japan\\
$^{2}$ Frontier Research Institute for Interdisciplinary Sciences, Tohoku University, Sendai 980-8578, Japan\\
$^{3}$ Astronomical Institute, Tohoku University, Sendai 980-8578, Japan\\
$^{4}$Department of Physics \& Astronomy, University of Nevada, Las Vegas, 4505 S. Maryland Pkwy, Las Vegas, NV 89154-4002, USA \\
}
\begin{document}

\date{Accepted ?; Received ??; in original form ???}

\pagerange{\pageref{firstpage}--\pageref{lastpage}} \pubyear{2008}

\maketitle

\label{firstpage}

%----------------------------------------------------------------------
%
% Abstract
%
%----------------------------------------------------------------------
\begin{abstract}
Star formation in high-redshift dwarf galaxies is a key to understand early galaxy evolution in the early Universe. 
Using the three-dimensional hydrodynamics code {\tt GIZMO}, we study the formation mechanism of cold, high-density gas clouds in interacting dwarf galaxies with halo masses of $\sim  3 \times 10^{7}~\Msun$, which are likely to be the formation sites of early star clusters. 
Our simulations can resolve both the structure of interstellar medium on small scales of $\lesssim 0.1$\,pc and the galactic disk simultaneously. 
We find that the cold gas clouds form in the post-shock region via thermal instability due to metal-line cooling, when the cooling time is shorter than the galactic dynamical time.
The mass function of cold clouds shows almost a power-law initially with an upper limit of thermally unstable scale. 
We find that some clouds merge into more massive ones with $\gtrsim 10^{4}~\Msun$ within $\sim 2~\rm Myr$.
Only the massive cold clouds with $\gtrsim 10^{3}~\Msun$ can keep collapsing due to gravitational instability, 
resulting in the formation of star clusters.  
In addition, we investigate the dependence of cloud mass function on metallicity and $\HM$ abundance,  and 
show that the cases with low metallicities ($\lesssim 10^{-2}~\Zsun$) or high $\HM$ abundance ($\gtrsim 10^{-3}$) cannot form massive cold clouds with $\gtrsim 10^{3}~\Msun$. 
\end{abstract}

%----------------------------------------------------------------------
%
% Keywords
%
%----------------------------------------------------------------------

%%%%%%%%%%%%%%%%%%%%%%%%%%%%%%%%%%%%%%%%%

%----------------------------------------------------------------------
%
% Section 1: Introduction
%
%----------------------------------------------------------------------

\section{Introduction}
\label{intro}
Understanding galaxy formation in the early universe is one of the major goals in today's astronomy.
Current structure formation theory based on the cold dark matter (CDM) model suggests that, at first,
galaxies form in low-mass haloes, i.e., dwarf galaxies, and then later evolve to massive galaxies as the halo mass grows via the merging process. 
In addition, high-redshift dwarf galaxies are the primary candidate sources of ionizing photons for reionising the Universe (e.g., \citealt{Yajima11, Jaacks13, Yajima14, Wise14, Kimm14, Ma15}).
Therefore, high-redshift dwarf galaxies are the key objects in understanding the early Universe. 
Upcoming next generation telescopes, such as the James Webb Space Telescope ({\it JWST}) and the Thirty Meter Telescope ({\it TMT}), aim at detecting these star-forming dwarf galaxies in the early Universe. 

The formation and evolution of high-redshift dwarf galaxies are significantly affected by 
star formation and stellar feedback, because the gas in these galaxies can easily be pushed out  
from the shallow gravitational potential well of the low-mass haloes due to feedback 
(e.g., \citealt{Wise12a, Johnson13, Hasegawa13, Hopkins14, Kimm14, Yajima17}).
Star formation occurs in cold gas clumps in galaxies. 
However, even the state-of-the-art cosmological simulations suffer from resolving 
the details of star formation and feedback processes because of the limitation in numerical resolution.  
Therefore, the details of star formation process in dwarf galaxies is still unclear. 

In this work, we investigate the formation of cold gas clumps in dwarf galaxies 
using high-resolution hydrodynamic simulations that are able to resolve the internal structure of ISM well. 
In particular, we focus on the major merger of dwarf galaxies that frequently occurs in the early Universe (e.g., \citealt{Genel09}). 
Recently, cosmological simulations of \citet{Katz15} showed that the major merger of dwarf galaxies induced the formation of massive gas clumps at the galactic center due to angular momentum transport by the tidal force or disk instability \citep[see also,][]{Devecchi12}. 
They suggested that the massive gas clump formed a compact star cluster, resulting in the formation of 
an intermediate-mass black hole via core-collapse of the star cluster (see also, \citealt{Yajima16}).
\citet{Trenti15} suggested that the interacting dwarf galaxies at high redshift were the likely sites of globular cluster  formation, and they were able to reproduce the local observations by combining cosmological $N$-body simulations and a semi-analytic model. 
Recent high-resolution cosmological simulations of \cite{Ricotti16} showed that
the gas clouds and star clusters formed in high-redshift dwarf galaxies, 
and some of them became very compact at the galactic center in major mergers of galaxies. 
In addition, cosmological simulations of \citet{Kim17} showed that the interacting dwarf galaxies 
formed massive compact star clusters with low metallicities that were similar to the local, observed globular clusters. 
Thus, most of recent cosmological simulations suggest that high-redshift interacting dwarf galaxies are likely to form massive gas clumps and star clusters at the galactic center
via rapid gas inflow. 
 
In the observations of local galaxies, it is well known that the formation of star clusters in
galactic disks can be enhanced due to galaxy mergers (e.g., \citealt{Alonso-Herrero00, deGrijs03}). 
These observations also suggest that gas clumps and star clusters may form in the galactic disks of interacting galaxies, and motivate the theoretical studies of their formation mechanisms. 
For example, \cite{IO15} investigated the thermal evolution of idealized colliding flow of a low metallicity gas using mesh-based three-dimensional hydrodynamic simulations, and showed that the cold gas clumps can form in the post-shock regions due to thermal instability (see also, \citealt{KI02}).  
They found that the condition for thermal instability is satisfied  
if the UV radiation field is strong enough to dissociate the $\HM$ molecules significantly ($\un{G}{0}> 10^{-3}$),
and if the metal-cooling is effective ($Z>10^{-3}\Zsun$) which is dominated by the $\rm{[C_{I\hspace{-0.1em}I}]}$ fine-structure line emission.
Therefore, they concluded that it is possible to form cold and dense gas clouds 
via thermal instability in interacting, low-metallicity galaxies under the UV irradiation.

On the other hand, the evolution of post-shock gas can also be affected by galactic dynamics, 
e.g., rotation of galactic disks, gravitational potential of haloes, and tidal forces from companion galaxies. 
However, the evolution of ISM at the post-shock region under the influence of galactic dynamics
has not been studied yet, especially in the context of high-redshift galaxy mergers.  
Given these situations, in the present work, we study the evolution of ISM in interacting galaxies 
using three-dimensional hydrodynamic simulations of merging galaxies with different metallicities and $\HM$ abundance. 

Our paper is organized as follows.
We describe our simulations and initial setup in Section~\ref{sec:method}. 
In Section~\ref{sec:result}, we present 
the formation of cold gas clumps, and its dependence on metallicity and $\HM$ abundance. 
In addition, we study the impact of different orbital parameters of galaxy merger on the formation of gas clumps. 
Finally, we summarize our main conclusions in Section~\ref{sec:discussion}.
%in prep. \\[5.0cm]

%%%%%%%%%%%%%%%%%%%%%%%%%%%%%%%%%%%%%%%%%

%----------------------------------------------------------------------
%
% Section 2:  Model and Method
%
%----------------------------------------------------------------------
\section{Model \& Method}
\label{sec:method}
\subsection{Simulation Setup}
In order to investigate the physical properties of ISM in interacting dwarf galaxies, here we 
use the Lagrangian hydrodynamics code $\tt{GIZMO}$
with the $\tt{Meshless\ Finite\ Mass}$ (MFM) method \citep{Hopkins15}. 
Using the initial condition generator DICE \citep{Perret14}, 
we construct disk galaxies with a virial mass of $\un{M}{vir}=3\times 10^{7}\,\Msun$ assuming a redshift $z=10$, and the virial temperature of the halo is $\Tvir = 8750\ \rm{K}$. 
The density profile of dark matter halo follows the NFW profile \citep{Navarro95}
with a concentration parameter $c=9.0$ and a spin parameter $\lambda=0.04$.
Note that, in this initial condition generator code, the virial radius of the halo is automatically adjusted for the assumed redshift, therefore the gas in halo is denser at a higher redshift than at $z=0$. 
The relevant cosmological parameters used are: $(\Omega_m, \Omega_\Lambda, h) = (0.3, 0.7, 0.71)$. 

The rotationally supported disk is composed of gas alone, and the disk mass is 10 per cent of the virial mass, i.e., $\un{M}{disk}=3\times 10^{6}\,\Msun$.
We assume an exponentially declining profile with a radial scale length $\un{h}{r}=0.1~\rm{kpc}$.
The mass resolution is $27~\Msun$ for dark matter and $0.3~\Msun$ for gas particles. 
The softening length of dark matter particles is set to $3.0\ \rm{pc}$. %\adr{(in physical scale?)}.
We employ adaptive softening lengths for gas particles, which is set to the same value as the smoothing length, 
and the minimum softening is set to $5.2\times 10^{-2}\ \rm{pc}$. 
Our simulations follow the ISM dynamics in the range  $1 ~\lesssim n_{\rm H}~{\rm cm^{-3}} \lesssim 10^{4}$, where $n_{\rm H}$ is the hydrogen number density.
In this density range, our numerical resolution satisfies the criterion to avoid artificial fragmentation by self-gravity \citep{Bate97,Susa08}.
In this work, we do not follow star formation, assuming that the galaxies are exposed to the external UV background \citep{HM12}.   
 
We put two identical galaxies on an orbital plane with inclinations $(\un{\phi}{1},\un{\phi}{2},\un{\theta}{1},\un{\theta}{2})$,
an impact parameter $(b)$, and an initial distance $(d=1.2\ \rm{kpc})$.
Once we start the simulation, galaxies collide with about their virial velocities ($\sim 10.8\ \rm{km\ s^{-1}}$).
Thus gas is heated up to almost virial temperature by shock compression.
The fiducial runs are calculated with $Z=0.1\Zsun$,
$b=0.3\ \rm{kpc}$, and  the inclination angles $\un{\theta}{1}=\un{\theta}{2}=\un{\phi}{1}=\un{\phi}{2}=0^{\circ}$. 
\subsection{Net Cooling and Thermal Instability}
\label{sec:cooling}
\begin{figure}
\begin{center}
\includegraphics[width=\columnwidth,bb=0 0 520 504]{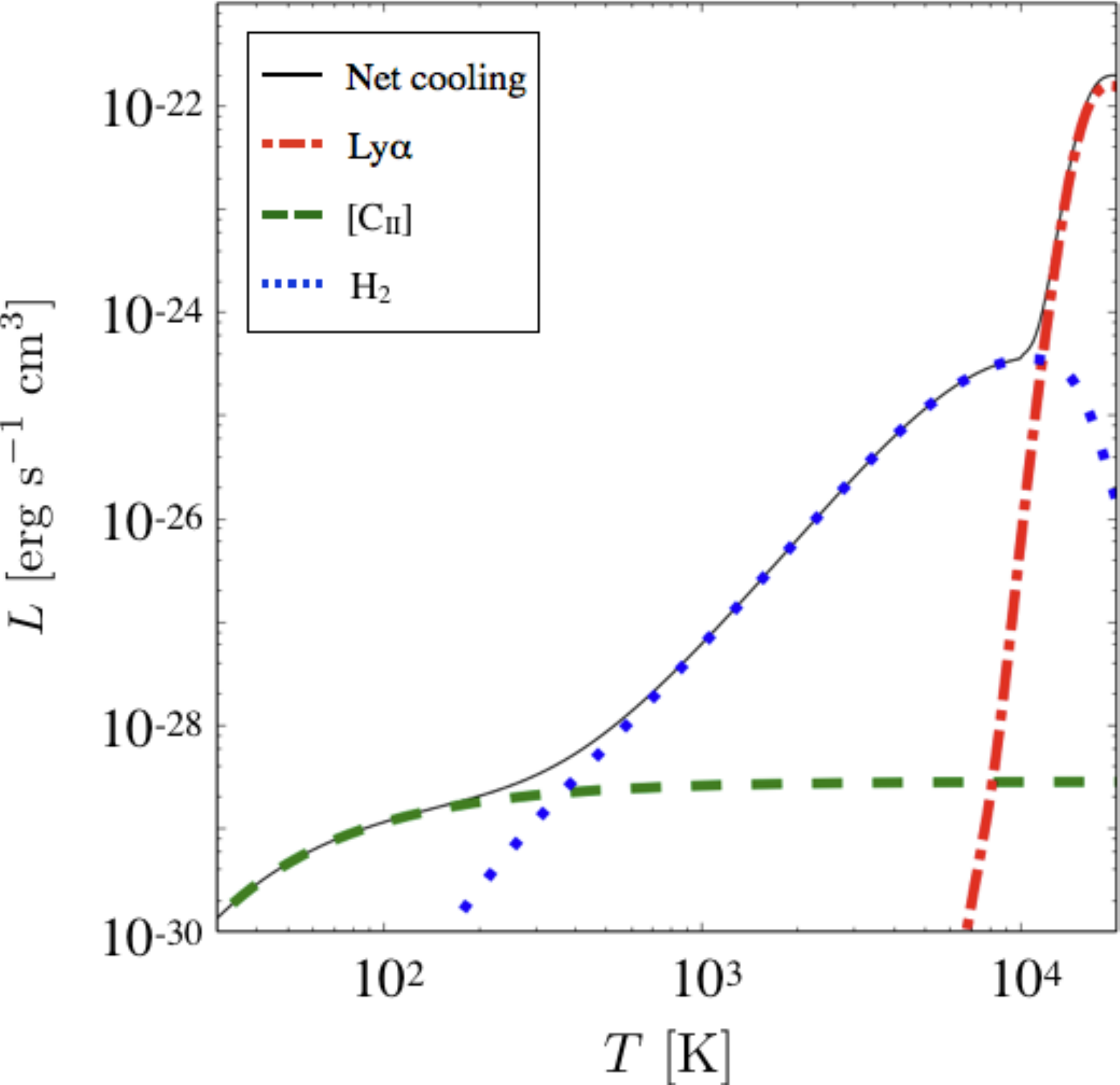}
\caption{
Radiative cooling rate as a function of gas temperature for $\lya$ line (red dot-dashed),  $\rm{[C_{I\hspace{-0.1em}I}]}$ line (green dashed; from Vazquez-Semadeni et al. 2007),  and $\HM$ line (blue dotted; from Galli \& Palla 1998).
Here we assume $Z=0.1\Zsun$ and $\un{f}{\HM}=10^{-3}$.
We include the $\HM$ cooling only in Sec.~\ref{sec:h2depend}.
}
\label{fig:cooling}
\end{center}
\end{figure}

We consider the heating and cooling at $T<10^{4}\ \rm{K}$ by 
taking into account the photoelectric heating of dust ($\Gamma$) and the $\rm{[C_{I\hspace{-0.1em}I}]}$ radiative cooling ($L(T)$), as proposed by \cite{KI02} and modified by \cite{Vazquez-Semadeni2007}
(see Eqns. 3 \& 4, and Fig. 2 in their paper):
\begin{eqnarray}
\Gamma &=& 2.0\times 10^{-26}\ \rm{erg\ s^{-1}}, \\
\frac{L(T)}{\Gamma} &=& 1.4\times 10^{-2}\sqrt{T}\exp{\left( \frac{-92}{T} \right)}\ \rm{cm^{3}}.
\end{eqnarray}
%
%\begin{equation*}
%\Gamma = 2.0\times 10^{-26}\ [\rm{erg\ s^{-1}}],
%\end{equation*}
%\begin{equation*}
%\frac{L(T)}{\Gamma} = 1.4\times 10^{-2}\sqrt{T}\exp{\left( \frac{-92}{T} \right)}\ [\rm{cm^{3}}].
%\end{equation*}
%
In addition, the $\lya$ cooling is considered. 
The above cooling rates are estimated under the assumption of ionization equilibrium state. 
The minimum temperature is set to $\un{T}{min}=30\ \rm{K}$ considering the temperature floor by the cosmic microwave background at $z \sim 10$. 
In our fiducial simulation, we do not include the cooling by H$_2$ assuming that the galaxies are exposed to the UVB. 
At $z\gtrsim 7$, the UVB strength can sensitively depend on the environment
due to patchy ionization structure of the IGM \citep{Iliev06}.
We study the impact of H$_2$ on the thermal evolution of ISM in Section~\ref{sec:h2depend}.
Figure~\ref{fig:cooling} presents the cooling functions of $\lya$, $\rm{[C_{I\hspace{-0.1em}I}]}$, and $\HM$.

%%Fig.1

Observations of our Milky Way galaxy suggest that the multi-phase state of ISM is induced via thermal instability \citep{Field65,MO77}, which we discuss below. 
Cold clouds are formed due to compression by the thermal pressure of ambient gas
when the net cooling rate $\Lambda$ satisfies the following condition:
% \citep{Balbus95, KI00}:
\begin{equation}
\label{eq:condition}
\left( \frac{\partial\Lambda}{\partial T}\right)_{p} < 0,
\end{equation}
where $\Lambda= \un{n}{H}^{2} L(T)\ \rm{erg\ cm^{-3}\ s^{-1}}$.
The condensation of clouds due to thermal instability proceeds under the isobaric condition.
The scale of thermal instability is limited within the range of $\un{\ell}{F}< \lambda <\un{\ell}{ac}$,
where $\un{\ell}{F}$ is the Field length \citep{Field65}, 
and $\un{\ell}{ac}$ is the acoustic length.
The Field length is defined by the ratio between thermal conduction rate and cooling rate:
\begin{equation}
\un{\ell}{F} =  \sqrt{\frac{\kappa\,T}{\Lambda}},
\end{equation}
where $\kappa$ is the thermal conductivity.
The acoustic length is defined by the propagation distance of sound waves within cooling time:
\begin{equation}
\un{\ell}{ac} = \un{c}{s}\un{t}{cool}.
\end{equation}
Using a typical post-shock density of $n_{\rm H} = 10\ \cc$ and $T=8000\ \rm{K}$,
we roughly estimate the acoustic length to be
\begin{equation} 
\un{\ell}{ac} = 16\left(\frac{\un{n}{H}}{10\ \cc}\right)^{-1} \left(\frac{Z}{0.1\Zsun}\right)^{-1}\ \rm{pc}.
\end{equation}
This corresponds to a cloud mass of
\begin{equation}
\begin{split}
 \un{M}{max} &= \frac{4\pi}{3}\rho \left(\frac{\un{\ell}{ac}}{2}\right)^{3}\\
 &\sim 6.4 \times10^{2} 
 \left( \frac{\un{n}{H}}{10\ \cc} \right)^{-2}
 \left( \frac{Z}{0.1\Zsun} \right)^{-3}
 \Msun, 
 \end{split}
 \label{eq:estimate}
\end{equation}
which we call the maximum cloud mass, $\un{M}{max}$, in the following. 

In order to calculate the isobaric contraction of thermally unstable gas, 
the simulation has to resolve the pressure gradient on smaller scales than the thermally unstable scale. 
The resolution of our simulations can resolve the acoustic length, 
while it cannot resolve the Field length. 
Thus, the lower limit of the cloud mass function is regulated by the numerical resolution. 
Therefore, in this work, we investigate the massive end of the cloud mass function.

\citet{KI02} showed that the thermal instability was enhanced at the post-shock region where the pressure was high. 
In our simulations, the temperature of gas disks is $\sim 1600~\rm K$ initially, 
then heated up to $\sim 8000~\rm K$ by shock compression in the merger process. 
In this work, we activate the radiative cooling of gas particles only for those that have $T>8000~\rm{K}$. 
Other gas particles are in the adiabatic state. 
This treatment makes it easier to understand the formation of cold clouds via thermal instability. 

The cooling and heating rates of low-metallicity gas are simply estimated by 
multiplying the metallicity ($Z/\Zsun$) to the rates of the solar value. 
\cite{IO15} validated the modified cooling function by comparing with the calculation of non-equilibrium chemistry. 
Here we estimate the cooling time in the post-shock region with $T=8000\ \rm{K}$
as follows:
\begin{equation}
\label{eq:cooltime}
\un{t}{cool}=\frac{\frac{3}{2}\un{k}{B} T}{ \un{n}{H} L(T,Z)} 
\approx 2.1\ \left(\frac{\un{n}{H}}{10\ \cc} \right)^{-1}
\left( \frac{Z}{0.1\Zsun} \right)^{-1}\ \rm{Myr}.
\end{equation}
This time-scale can be longer than or comparable to the dynamical time-scales of dwarf galaxies,
such as the crossing time of merging dwarf galaxies, or the rotation period of galactic disk. 
Therefore, the gas dynamics of ISM is also affected by the tidal force and differential rotation. 
\subsection{Clump Finder}
\label{sec:clumpfinder}
The cold gas clouds in our simulation are identified by the friend-of-friend (FOF) method.
First, we identify the particles in the post-shock region with $T<500~\rm K$
and density greater than $50\times \un{\rho}{bg}$, where
$\un{\rho}{bg}$ is the mean density of warm background gas.
Next, we connect the two cold particles if their smoothing length is greater than their distance.
The minimum threshold number of gas particles for identifying a cold cloud is 100,
corresponding to $30\, \Msun$ as the minimum cloud mass ($\un{M}{min}$).

%%%%%%%%%%%%%%%%%%%%%%%%%%%%%%%%%%%%%%%

%----------------------------------------------------------------------
%
% Section 3:  Results
%
%----------------------------------------------------------------------

%Fig.2
\begin{figure*}
\begin{center}
\includegraphics[width=2.0\columnwidth, bb=0 0 881 219]{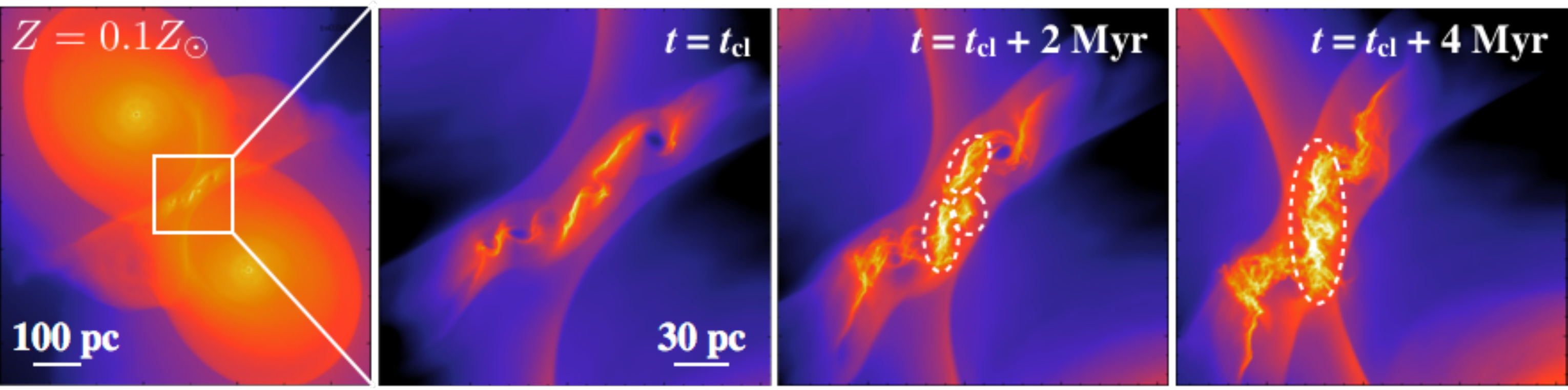}
\caption{
Time evolution of the column density around the shock-compressed region.
The left two panels show the epoch when our clump finder identifies cold clouds at first ($t=\un{t}{cl}$) in our fiducial run: $Z=0.1\Zsun,\ b=0.3\ \rm{kpc}$, prograde-prograde merger. 
Three marked cold clouds at $t=\un{t}{cl}+2~\rm{Myr}$ merge and form a very massive cloud ($\un{M}{cl}>10^{4}\Msun$) within $2~\rm{Myr}$ (see also Fig.~\ref{fig:massdist}).
}
\label{fig:timesnaps}
\end{center}
\end{figure*}

%Fig.3
\begin{figure}
\begin{center}
\includegraphics[width=0.95\columnwidth, bb=0 0 538 818]{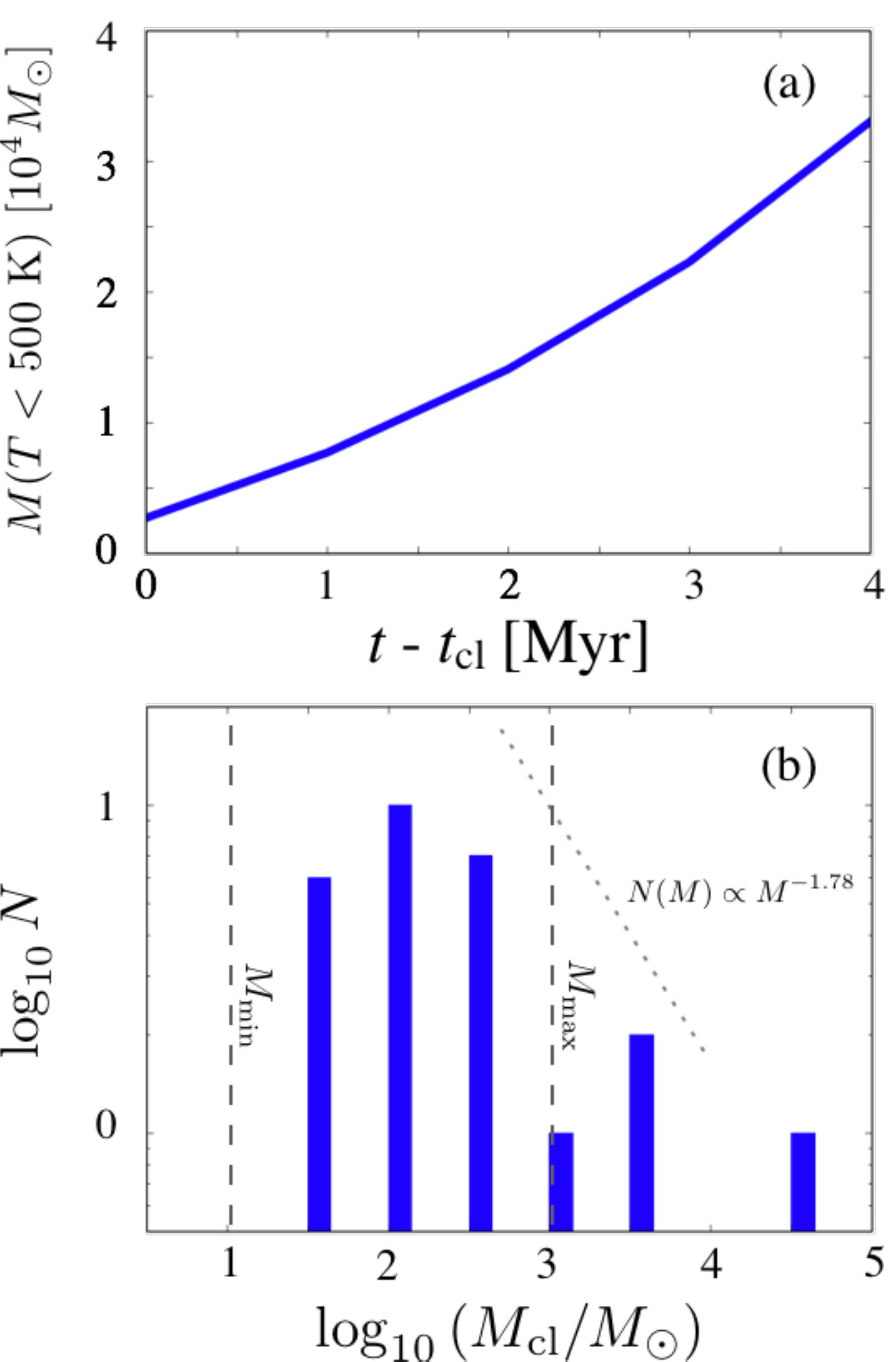}
\caption{
{\it Panel (a)}: Time evolution of the total mass of cold gas ($T< 500~\rm{K}$) in our fiducial run.
{\it Panel (b)}: Mass distribution of identified cold clouds in each case at $t=\un{t}{cl}+4~\rm{Myr}$.
Two-dotted lines represent the mass limits: $\un{M}{min}=30\Msun$, $\un{M}{max}=6.4\times 10^2\Msun$ (Eq.~\ref{eq:estimate}).
The power-law $N(M)\propto M^{-1.78}$ is shown as a reference (Eq.\,\ref{eq:massfcn}). 
}
\label{fig:massdist}
\end{center}
\end{figure}

\section{Results}
\label{sec:result}
\subsection{Formation of cold gas clouds with $Z=0.1\Zsun$
(fiducial run)}
\label{sec:fiducail}

First, we study the formation mechanism of the cold clouds in the fiducial run.
Figure~\ref{fig:timesnaps} shows the maps of gas column density at each snapshot. 
In the shocked region, the filamentary and clumpy structures form due to galaxy merger. 
Some clouds merge into more massive ones as marked by the white-dashed circles
in a short time-scale of $\lesssim 3~\rm Myr$.
Since the time-scale is shorter than the lifetime of massive stars, 
massive star clusters can form in the merged massive clouds.  

Figure~\ref{fig:massdist}(a) shows the time evolution of the total mass of cold gas clouds from the time when our clump finder identifies cold clouds at first ($\equiv \un{t}{cl}$).
The total cold gas mass increases rapidly via thermal instability from $\sim 0.3\times 10^4\,\Msun$ to $3.3\times 10^4\,\Msun$ within 4\,Myr,  with a mass accumulation rate of $d\un{M}{cl}/dt \sim 10^{-2}~\Msun~{\rm yr^{-1}}$. 

Figure~\ref{fig:massdist}(b) shows the mass distribution of cold clouds at $t=\un{t}{cl}+4~{\rm Myr}$.
As time proceeds, the number of clouds increases mainly in the mass range of $10^{2}-10^{3}~\Msun$.
As explained in Sec.~\ref{sec:method}, the scale of perturbations which can grow via thermal instability
is limited within $\un{\ell}{F}<\lambda<\un{\ell}{ac}$.  
The cloud mass function takes a power-law form, 
\begin{equation}
N(M)dM \propto M^{(\delta-3)/3-2} dM, 
\label{eq:massfcn}
\end{equation}
where $\delta$ is the power-law index of the three-dimensional power spectrum of the seed density fluctuations \citep{Hennebelle07,Hennebelle08}.
It gives $N(M)\propto M^{-1.78}$ for a Kolmogorov-type fluctuation of $\delta=11/3$, which we show in 
Figure~\ref{fig:massdist}(b) as a reference. 
Our result also shows a similar power-law-like cloud mass distribution close to $N(M) \propto M^{-2}$, which is a Press-Schechter type mass function.  
The minimum cloud mass is regulated by the threshold in our clump finder, and the vertical dashed lines represent the minimum and maximum cloud masses as derived in Eq.~\ref{eq:estimate}.
Most of the clouds distribute in the range $\un{M}{min}<\un{M}{cl}<\un{M}{max}$. 
However, unlike the results of \citet{IO15} who studied the idealized colliding flow, 
our result shows an extended tail at the massive end that exceeds $\un{M}{max}$.
This is due to the coalescence of some massive clouds; for example, the three clouds in Fig.~\ref{fig:timesnaps} marked by white dashed circles, merge within $2~\rm{Myr}$ in the tidal arm, resulting in a very massive cloud with $\un{M}{cl}=2.6\times 10^{4}~\Msun$.

%%Fig.4
\begin{figure*}
\begin{center}
\includegraphics[width=5in,bb=0 0 804 702]{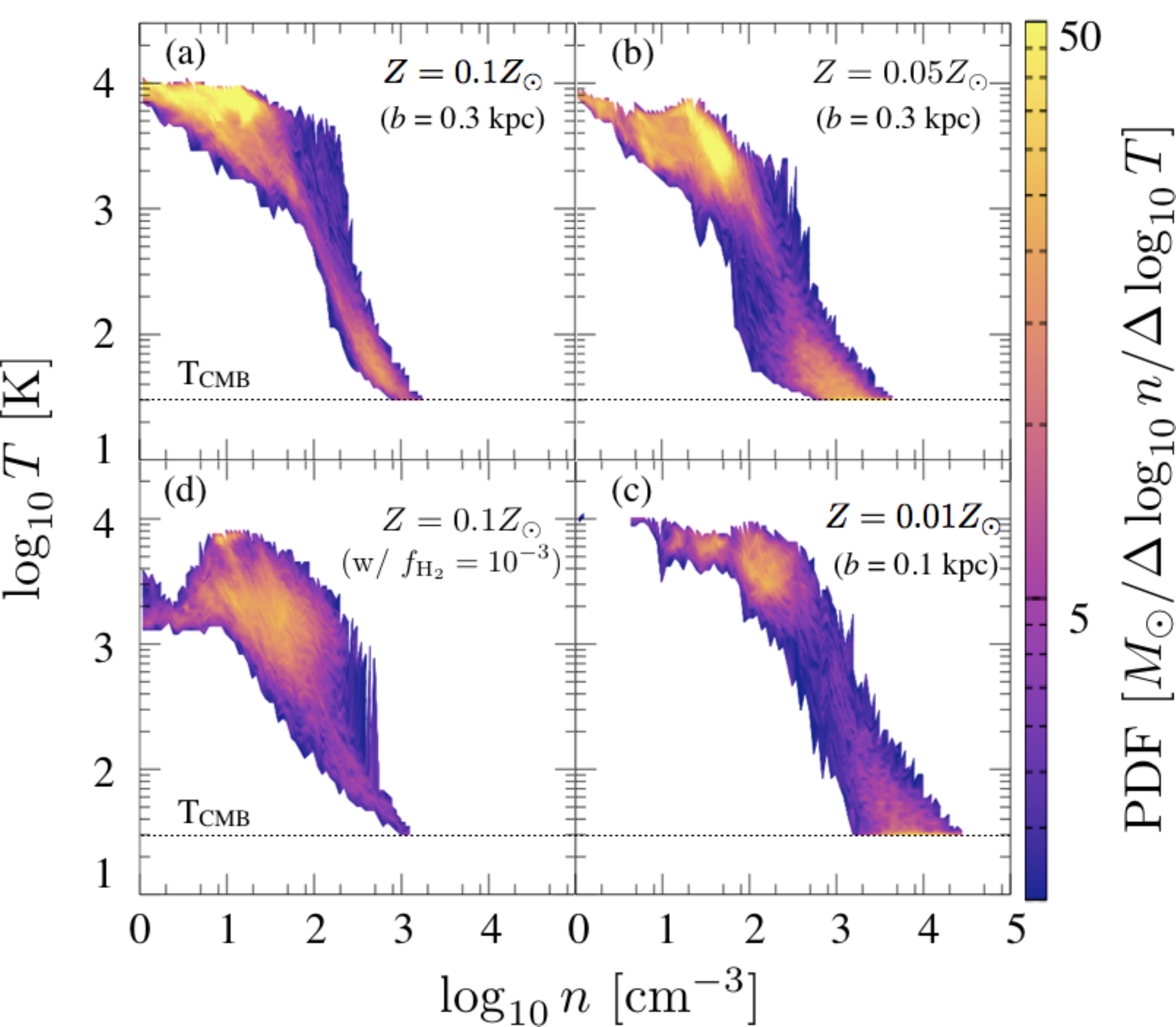}
\caption{
Mass-weighted phase-diagram of cold clouds identified by our clump finder
and its vicinities at $t=\un{t}{cl}+2~\rm{Myr}$. 
{\it Panel (a)}: fiducial run ($Z=0.1\Zsun,~b=0.3\ \rm{kpc}$); 
{\it Panel (b)}: lower metallicity case ($Z=0.05\Zsun,~b=0.3~\rm{kpc}$); 
{\it Panel (c)}: almost head-on merger with extremely low-metallicity case ($Z=0.01\Zsun,~b=0.1~\rm{kpc}$); 
{\it Panel (d)}: with a constant $\HM$ abundance ($Z=0.1\Zsun,~b=0.3~\rm{kpc},~\un{f}{H_2}=10^{-3}$).
The panels (a)-(c) show the two-phase structure via thermal instability.
In panel (d), however, the two-phase structure does not appear, because the $\HM$ cooling with a critical abundance breaks the condition for thermal instability (Eq.~\ref{eq:condition}).
Dotted lines represent the temperature floor of 30\,K by the CMB at $z \sim 10$.
}
\label{fig:phase}
\end{center}
\end{figure*}

As shown in recent high-resolution cosmological simulations, 
galaxy interaction can cause massive gas inflow towards the galactic center
and form a massive gas clump at the galactic center (e.g., \citealt{Katz15, Ricotti16}). 
Here we find that massive gas clouds could form even within the galactic disk 
via thermal instability and cloud mergers. 
Note that the high resolution with $< 1~\rm pc$ is required to follow the cloud formation via thermal instability 
as we discuss in the Appendix. 
Current cosmological simulations are still difficult to resolve such small scales (e.g., \citealt{Kim17, Yajima17}).
In addition,  the massive clouds could form within several Myr, 
while the mass inflow to the galactic center occurs over the galaxy dynamical time $> 10~\rm Myr$. 
Therefore, if massive star clusters form in the clouds in the galactic disk, 
their intense stellar feedback can evaporate most of the gas from the galaxy.
This can drastically change the star formation history of high-redshift galaxies. 
In our future work, we will study the impact of star formation and stellar feedback on galaxy evolution
with massive gas clump formation in galactic disks. 

Figure~\ref{fig:phase}(a) shows the mass-weighted phase-diagram of 
the most massive cloud and its vicinities in the fiducial run at $t=\un{t}{cl}+2\ \rm{Myr}$.
In the post-shock region, the gas separates into two-phase structure with 
warm--low density $(n,T)\approx (20\,\cc,\ 5000\,\rm{K})$ and
cold--high density $(500\,\cc,\ 50\,\rm{K})$.
This means that the perturbations in the warm phase contracts almost isobarically, 
leading to the formation of cold gas clouds, in pressure equilibrium with the warm gas.
These clouds are not formed via gravitational instability. 
The Jeans mass ($M_{\rm J}$) in the post-shock region is estimated as 
\begin{equation}
M_{\rm J} = 7.5 \times 10^{6} \left(  \frac{n_{\rm H}}{10~\rm cm^{-3}}\right)^{-\frac{1}{2}}
\left( \frac{T}{8000~\rm K} \right)^{\frac{3}{2}} \Msun.
\end{equation}

The masses of cold clouds in our simulation are much lower than this Jeans mass. 
%The masses of cold clouds in our simulation are lower than this Jeans mass. 
Note that the above Jeans mass is comparable to the total gas mass in the galaxy. 
Therefore, gravitational instability does not play a role in the formation of gas clumps in the galactic disks initially. 
As the density increases and the temperature decreases in the forming clouds, 
the Jeans mass decreases.  Then the massive cold clouds with $\un{M}{cl}>10^{3}\Msun$ alone exceed the Jeans criteria, and hence can collapse further by self-gravity. 
%Then all identified cold clouds with $\un{M}{cl}>15\Msun$ exceed the Jeans criteria, and hence can collapse further by self-gravity and form star clusters in the end. 
Subsequent gravitational collapse of the cold clouds cannot be followed in our simulations due to resolution limit.

%%%%%%%%%%%%%%%%%%%%%%%%%%%

\subsection{Metallicity dependence}
\label{sec:metaldepend}

The first galaxies are metal enriched by galactic outflows from nearby star-forming galaxies or type-II supernovae of Population III stars in mini-haloes that formed earlier. 
Therefore, the metallicity of first galaxies is likely to depend on the formation site (e.g., \citealt{Maio11, Wise12a, Smith16}). 
In this section, we study the metallicity dependence of cold cloud formation
in the spirit of investigating the environmental effect. 

\begin{figure}
\begin{center}
\includegraphics[width=\columnwidth, bb=0 0 615 1321]{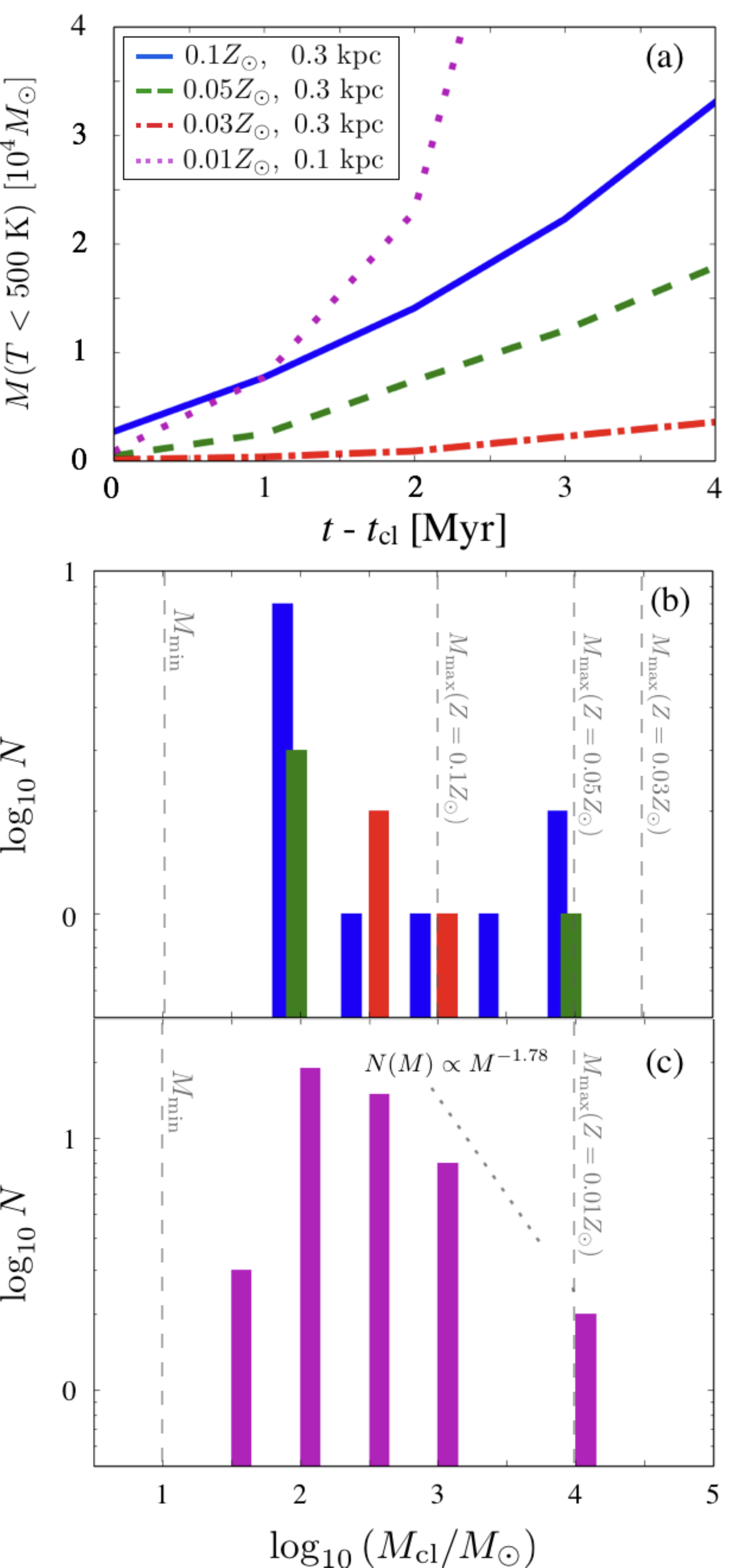}
\caption{
Metallicity dependence of cold cloud formation.
{\it Panel (a)}:  Time evolution of the cold gas mass in the weakly interacting galaxies ($b=0.3~\rm{kpc}$) with low metallicities, $Z=0.1\Zsun$ (blue), $0.05\Zsun$ (green), and $0.03\Zsun$ (red), 
and in the close encounter case ($b=0.1$\,kpc) with extremely low metallicity $Z=0.01\Zsun$ (purple).
{\it Panels (b) \& (c)}:  Mass function of identified cold clouds in the case of $b=0.3$\,kpc and $b=0.1$\,kpc at $t=\un{t}{cl}+2\ \rm{Myr}$, respectively. 
The power-law $N(M)\propto M^{-1.78}$ is shown as a reference (Eq.\,\ref{eq:massfcn}). 
} 
\label{fig:metal_depend}
\end{center}
\end{figure}

We vary the metallicity from $Z=0.1\Zsun$ to $0.01\Zsun$ 
under the same condition as the case with $b=0.3\ \rm{kpc}$. 
The typical density of post-shock region in this condition is $n_{\rm H} \sim 10~\rm cm^{-3}$. 
Figure~\ref{fig:metal_depend} presents the mass distribution of cold clouds with different metallicities at $t=t_{\rm cl}+2~\rm Myr$. 
We find that the simulations with higher metallicities than $\sim 0.03~\Zsun$ successfully form cold clouds. 
Because the maximum cloud mass depends on metallicity strongly ($M_{\rm max} \propto Z^{-3}$; Eq.~\ref{eq:estimate}), we can expect that more massive clouds appear as metallicity becomes lower.
From the simulation results, however, we find that the mass of the most massive cloud
in the simulation with $Z=0.03 \Zsun$ is much lower than Eq.~(\ref{eq:estimate}).
This is related to the cooling and galactic dynamical time scales. 
The galactic dynamical time is roughly estimated as
\begin{equation}
t_{\rm dyn} \sim \frac{R_{\rm vir}}{V_{\rm c}} = 28.4\ {\rm Myr}~\left(\frac{1+z}{11}\right)^{-3/2},
%$t_{\rm dyn} \sim \frac{R_{\rm vir}}{V_{\rm c}} = 28.4\ {\rm Myr}~\left(\frac{1+z}{11}\right)^{-3/2}$,
\end{equation}
where $R_{\rm vir}$ is the virial radius and $V_{\rm c}$ is the halo circular velocity ($\approx$ colliding velocities in our simulations).
As the galactic dynamical time becomes close to the cooling time, 
the strong tidal force can regulate the growth of large-scale perturbations. 
In the simulations with $Z\leq0.02\Zsun$, cold clouds do not appear in the post-shock region.
Therefore we suggest that cold clouds are unlikely to form in merging high-$z$ dwarf galaxies
with $Z \lesssim 10^{-2}~\Zsun$ and $\un{n}{H}\lesssim 10~\rm cm^{-3}$ in the post-shock region. 
Then, some fraction of cold gas can be converted into star clusters, although the conversion efficiency is not understood well (e.g., \citealt{Geen17}). 
Therefore, our results suggest that stellar distribution can sensitively depend on the metallicity of galaxies. 
Galaxies with lower metallicity than the critical value tend to form stars at the galactic center, 
while other galaxies can have stellar disks.  

Note that the density at the post-shock region can be increased by adopting different collision parameters for the merging galaxies. 
This reduces the critical metallicity for the cold cloud formation, which will be discussed in the next subsection.

%%Fig.5

%%%%%%%%%%%%%%%%%%%%%%%%%%%%%%%%%%

\subsection{Different orbital parameters}
\label{sec:collisiondepend}

In the growth of galaxies via major merger process, 
various orbital parameters and inclination angles can be considered (e.g., Khochfar \& Silk 2006). 
In this section, we investigate the impact of different parameters on the formation of cold gas clouds. 

As discussed in Sec. \ref{sec:metaldepend}, the low-metallicity case of $Z=0.01~\Zsun$ with $b=0.3~\rm{kpc}$
failed to form cold clouds due to the longer cooling time than the galaxy dynamical time. 
If the gas density at the post-shock region increases, the cooling time becomes shorter, and it is likely to 
cause the formation of cold clouds even with such a low-metallicity. 
Thus we study a case with a smaller impact parameter of $b=0.1~\rm{kpc}$ and $Z=0.01\Zsun$, 
which is likely to give higher gas density in the post-shock region. 

Figure~\ref{fig:metal_depend} presents the time evolution of the total mass of cold gas and the cloud mass distribution. 
In this case, cold clouds successfully form, unlike the case with $b=0.3~\rm kpc$. 
The total mass of cold clouds is $ 7.5 \times 10^{4}~\Msun$ at $t=t_{\rm cl}+3~\rm Myr$. 
The clump mass function has a peak at $M_{\rm cl} \sim 100~\Msun$, and is consistent with the power-law $N(M) \propto M^{-1.78}$ at the higher mass side (see Eq.\,\ref{eq:massfcn}). 

The corresponding phase-diagram of gas in the post-shock region was shown in Fig.~\ref{fig:phase}(c). 
The density of warm gas is $\un{n}{H}\approx100~\cc$ in the case of closer collision, 
and the gas separates into two phases due to thermal instability. 

Note that, gas clouds can be optically thick to the $\rm{[C_{I\hspace{-0.1em}I}]}$ emission, 
when their density becomes higher than $\sim 10^{3}~\cc$ \citep{Draine11}. 
For the high-density gas,  our simulations assuming optically-thin are likely to overestimate the cooling rate. 
Therefore, the cold gas cannot collapse further more at a specific gas density, 
when the cooling time becomes longer than the free-fall time. 
\cite{Omukai00} introduced the escape probability
to any line cooling processes based on the assumption that
the velocity was proportional to the radius from the center of cloud
(see also, \citealt{Takahashi83}).
We will incorporate the escape probability of the line cooling into our code in the future work. 

\begin{figure}
\begin{center}
\includegraphics[width=\columnwidth, bb=0 0 604 951]{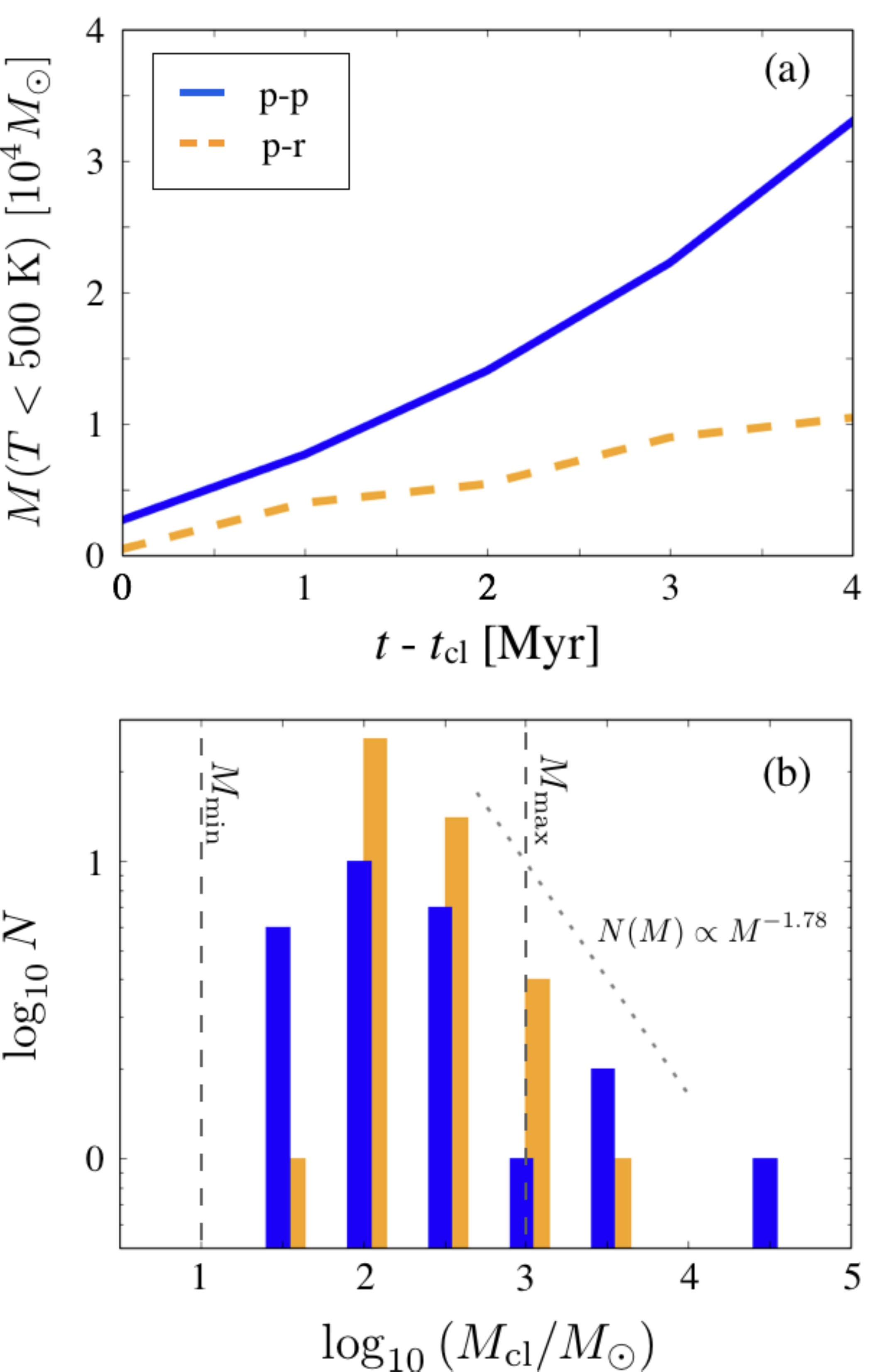}
\caption{
Same as Figure~\ref{fig:metal_depend}, but for different inclination angles. 
Blue and orange colors represent the prograde--prograde merger (fiducial run, $\un{\theta}{1}=\un{\theta}{2}=0^{\circ}$) and the retrograde--prograde merger ($\un{\theta}{1}=0^{\circ}, \un{\theta}{2}=180^{\circ}$), respectively.
The power-law $N(M)\propto M^{-1.78}$ is shown as a reference (Eq.\,\ref{eq:massfcn}).  
} 
\label{fig:incli_depend}
\end{center}
\end{figure}

\begin{figure}
\begin{center}
\includegraphics[width=\columnwidth, bb=0 0 879 464]{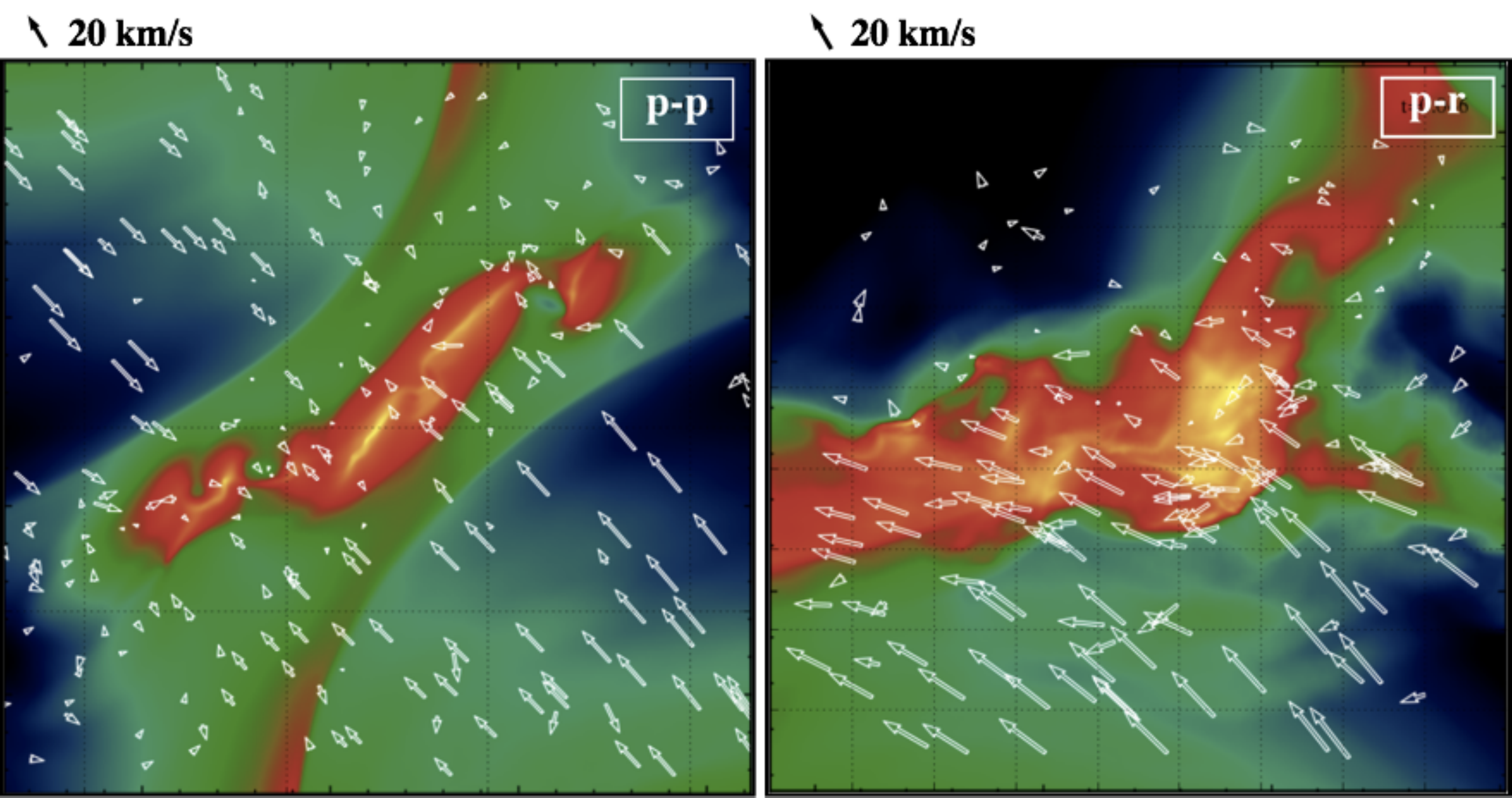}
\caption{
Velocity fields are shown by arrows together with gas column density. 
The length of white arrows is proportional to the absolute velocity values. 
Left and right panels show the snap shots at $t=t_{\rm cl}-1~\rm Myr$ in 
the prograde-prograde and the prograde-retrograde runs, respectively. 
} 
\label{fig:vfield}
\end{center}
\end{figure}

Next we study the case with retrograde--prograde merger, i.e., the inclination angle is $180^{\circ}$ flipped. 
Other parameters are the same as those in the fiducial run.
The time evolution of cold gas mass and the mass distribution are presented in Fig.~\ref{fig:incli_depend}.
We find that, in the retrograde--prograde merger, the final cold mass is three times less than that of the fiducial case.
This difference can come from the suppression of clump formation due to a shear flow. 
Figure~\ref{fig:vfield} presents the velocity field with arrows overlaid on the projected gas distribution. 
In the case of prograde--prograde merger, the velocity becomes small at the post-shock region. 
On the other hand, the retrograde--prograde merger induces an oblique shock. 
In this case, the parallel component of the velocity to the direction of galaxy collision can remain, resulting in the shear flow as seen in the figure. 

\begin{figure}
\begin{center}
\includegraphics[width=\columnwidth, bb=0 0 665 521]{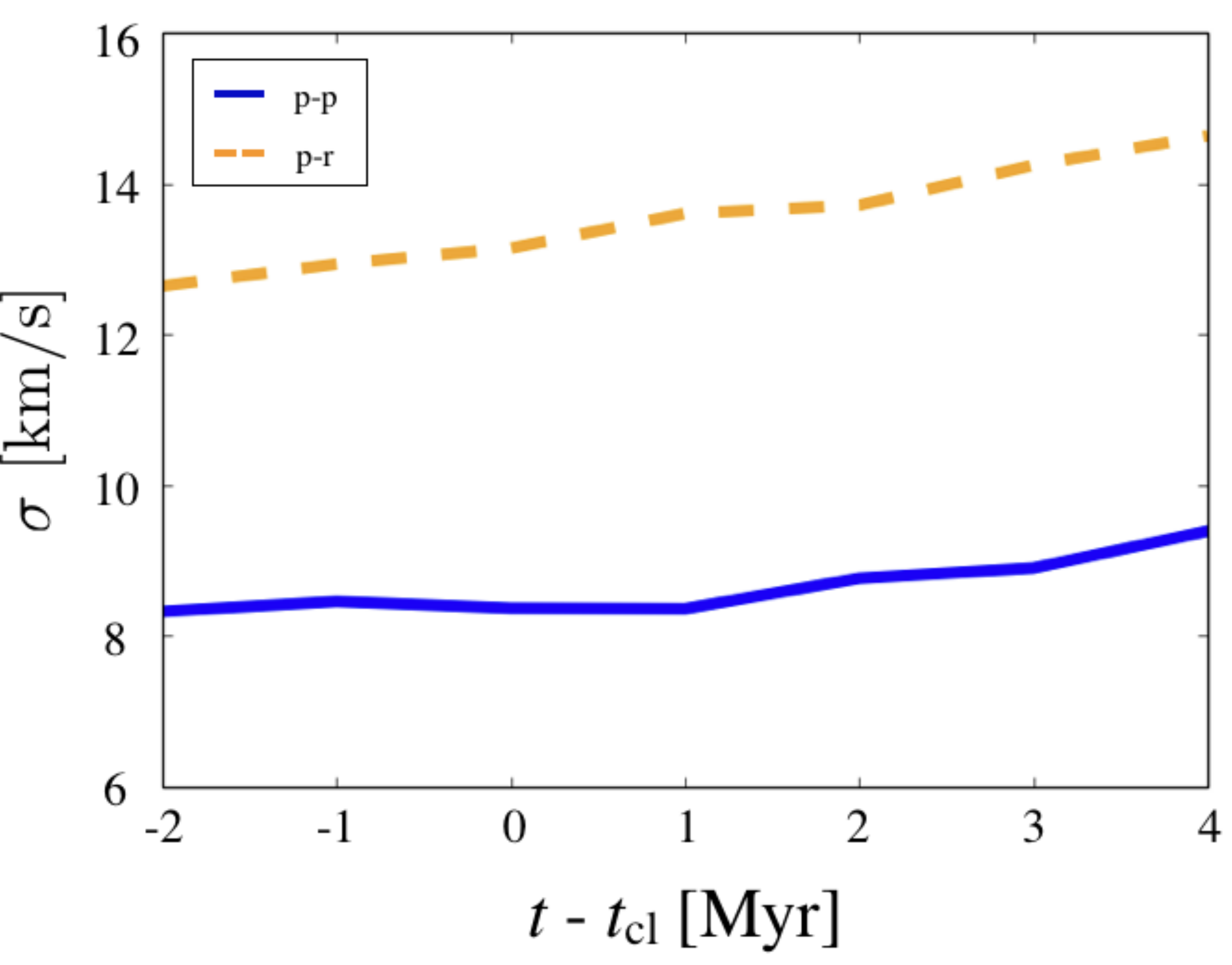}
\caption{
Time evolution of velocity dispersion in the post-shock regions. 
Blue-solid and orange-dashed lines show the prograde--prograde (p--p) 
and the prograde--retrograde (p--r) runs, respectively. 
}
\label{fig:veldisp}
\end{center}
\end{figure}

The time evolution of the velocity dispersion at the post-shock region is presented in Figure~\ref{fig:veldisp}.
Unlike the prograde--prograde merger, the retrograde--prograde merger keeps the velocity dispersion larger than $\sim 12~\rm km~s^{-1}$, which is higher than the sound speed at the post-shock region $c_{\rm s} \sim 7-8~\rm km~s^{-1}$. 
Therefore, the shear flow can disturb the compression of gas by the isotropic thermal pressure of ambient gas. 

Given the results presented in this subsection, we suggest that the orbital motion and inclination angle are important parameters to control the clump formation via thermal instability, which could not be investigated in previous works using idealized colliding flows (e.g., \citealt{KI02, IO15}). 
In our future work, we will study the statistical properties of cold gas clumps in merging galaxies 
taking the impact parameters and inclination angles from cosmological simulations.

%%Fig.6

%%Fig.7

%%Fig.8

%%%%%%%%%%%%%%%%%%%%%%%%%%%%%%%%%%%%%

\subsection{Impact of hydrogen molecules}
\label{sec:h2depend}

In the previous subsections, we assumed that $\HM$ is completely dissociated by the Lyman-Werner band radiation from nearby star-forming galaxies.
However if the UV radiation field is not so strong,
$\HM$ can form and contribute significantly to the radiative cooling at $T<10^{4}$\,K.
In order to study the impact of $\HM$ cooling on cloud formation, 
we carry out additional simulations with $\HM$ under the assumption of constant $\HM$ abundance
of $\un{f}{\HM} \equiv \un{n}{\HM}/\un{n}{H} =10^{-5},\ 10^{-4}~\&~10^{-3}$.
We incorporate the cooling function of $\HM$ from \citet{Galli98}
into our cooling rates. 
The orbital parameters and halo mass of the simulation are the same as the fiducial one. 

\begin{figure}
\begin{center}
\includegraphics[width=\columnwidth,bb=0 0 600 917]{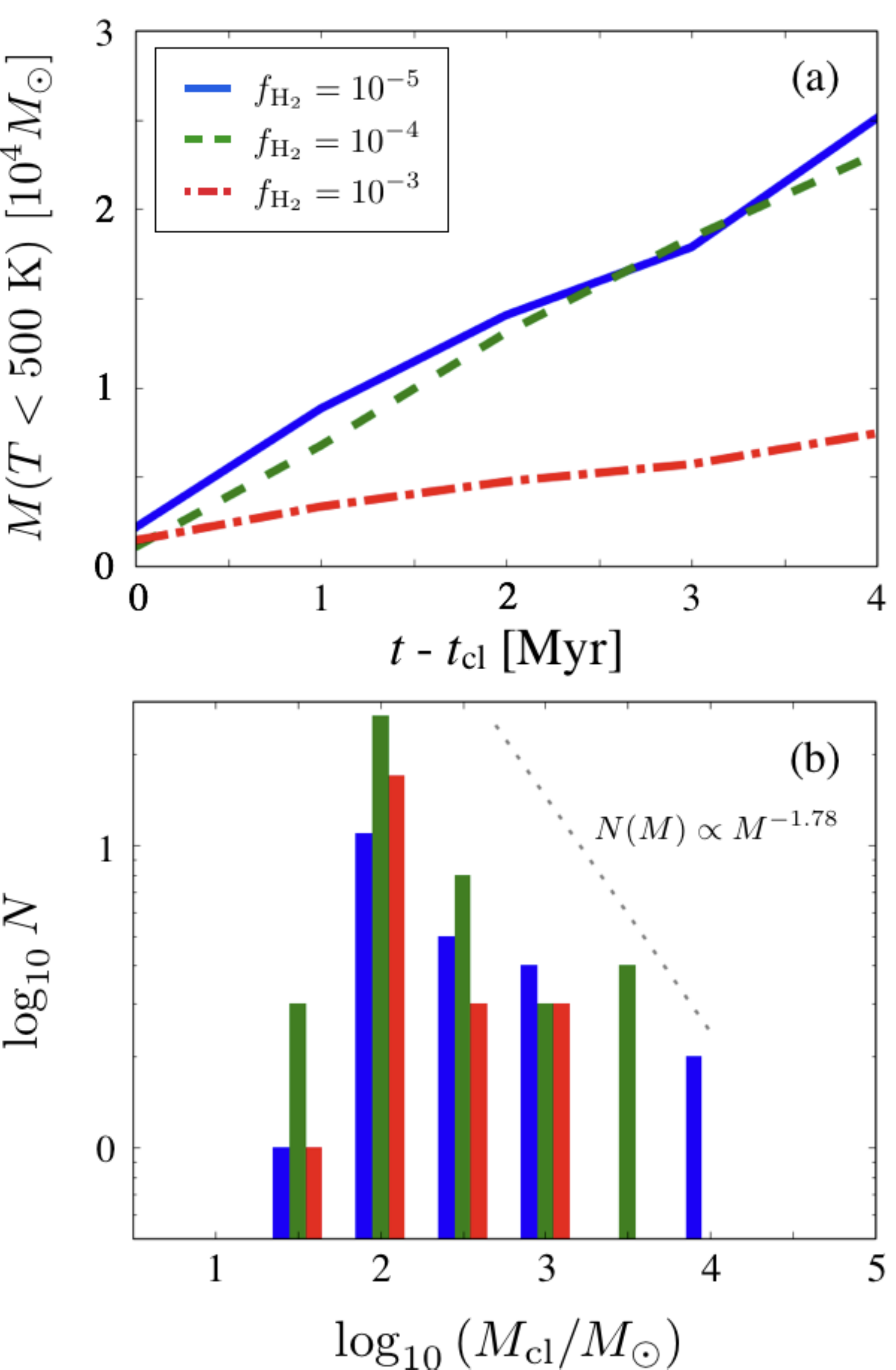}
\caption{
%Hydrogen molecules dependence of cold cloud formation.
Formation of cold clouds under different abundances of hydrogen molecules.
Panel (a): Time evolution of cold gas mass in weak interacting low-metal galaxies ($b=0.3~\rm{kpc}$, $Z=0.1\Zsun$) including some $\HM$ abundances $\un{f}{\HM}=10^{-5}$ (blue), $10^{-4}$ (green), $10^{-3}$ (red).
Panel (b):  Mass function of identified cold clouds in each case.
The power-law $N(M)\propto M^{-1.78}$ is shown as a reference (Eq.\,\ref{eq:massfcn}).  
} 
\label{fig:hm_depend}
\end{center}
\end{figure}

Figure~\ref{fig:hm_depend}(a) shows the time evolution of total mass of cold gas ($T \leq 500\ \rm{K}$).
There is almost no difference between $\un{f}{\HM}=10^{-5}$ and $10^{-4}$ cases,
while in the $\un{f}{\HM}=10^{-3}$ case the mass of cold gas is clearly less.
This is simply because the efficient $\HM$ cooling breaks the condition of the net cooling rate for thermal instability (see, Eq. 1).

Figure~\ref{fig:phase}(d) presents the phase-diagram of the post-shock region in the simulation with $\un{f}{\HM}=10^{-3}$. 
We find that the formation of multi-phase ISM is suppressed and  
most of the gas particles have $T \sim 1000~\rm K$ due to $\HM$ cooling. 
The critical $\HM$ abundance ($f_{\rm \HM, crit}$) for inducing the thermal instability sensitively depends on the metallicity. 
In our current simulation, it implies 
\begin{equation}
f_{\rm \HM, crit} \sim 10^{-3} \left( \frac{Z}{0.1~\Zsun} \right). 
\label{eq:Hcrit}
\end{equation}

Figure~\ref{fig:hm_depend}(b) shows the mass distribution of cold clouds in each simulation.
%Some cold clouds formed via the thermal instability ($\un{f}{2}=10^{-5},10^{-4}$) are massive enough to collapse gravitationally because their masses exceed the Jeans mass $5.1\times 10^{2}\Msun$ at $\un{n}{H}= 500~\cc$,\ $T=50~{\rm{K}}$, while most of clouds formed in $\un{f}{2}=10^{-3}$ case are gravitationally stable.
In the cases of $\un{f}{\HM}=10^{-5}$ \& $10^{-4}$, some cold clouds formed via thermal instability, 
and they exceed the Jeans mass $M_{\rm J} \sim 5.2\times 10^{2}~\Msun$ for $\un{n}{H}= 500~\cc$~and~$T=50~{\rm{K}}$.
Therefore they will collapse gravitationally, and are likely to be the formation sites of massive star clusters. 
On the other hand, in the case of $\un{f}{\HM}=10^{-3}$, massive cold clouds do not form. 
Thus, we suggest that $\HM$ plays a vital role in the formation of massive star clusters in interacting dwarf galaxies at high redshift. 

In addition, our results suggest that star formation and stellar distribution can sensitively depend on 
the formation sites of galaxies. At high redshifts ($z \gtrsim 6$), the UV background is likely to be inhomogeneous \citep{Johnson14, Agarwal14}, which will cause different H$_2$ abundance in different galaxies depending on their location. 
If galaxies are clustered, star clusters can efficiently form in galactic disks with little H$_2$ as we showed earlier. 
On the other hand, in galaxies far from other star-forming galaxies, 
star clusters will form via other pathways, e.g., Toomre instability in cold, high-density disks due to $\HM$ cooling, or gas inflow to galactic centers. 
For example, \cite{Susa04} showed that the morphology of galaxies could change significantly in different UV background fields due to the suppression of star formation. 
\cite{Okamoto08} indicated that the UV background could reduce the cosmic star formation history significantly because of the suppression of star formation in dwarf galaxies 
(see also, \citealt{Hasegawa13, Yajima17}). 
These previous works focused on the photo-evaporation effects of gas by 
the Lyman-continuum photons. 
Our results are complementary to their work, and we show that the Lyman-Werner photons can also have large impact on star formation and stellar structure of high-$z$ galaxies. 
The above effects on the statistical natures of high-$z$ dwarf galaxies should be investigated with cosmological simulation with inhomogeneous UV background. 

%----------------------------------------------------------------------
%
% Section 4:  Discussion
%
%----------------------------------------------------------------------

\section{Summary and Discussion} 
\label{sec:discussion}

%\subsection{}
%\label{sec:scaling}
We investigate the formation of cold, high-density clouds in merging dwarf galaxies using high-resolution three-dimensional hydrodynamical simulations.
Our simulations follow the thermal and dynamical evolution of ISM in low-mass galaxies with a halo mass of $\un{M}{vir}\sim 3\times 10^{7}~\Msun$ at $z=10$.   %, corresponding to the virial $\un{T}{vir}\sim8000~\rm{K}$).  
The virial size and density of the halo was set up appropriately for $z=10$ in the initial condition. 
Using various conditions with different metallicities, $\HM$ abundance, and orbital parameters, 
we study the impact of these quantities on the formation efficiency of cold clouds. 

Our major findings are as follows:
\begin{itemize}
	\item The shock compressed gas becomes thermally unstable via $\rm{[C_{I\hspace{-0.1em}I}]}$ cooling, and the cold clouds successfully form only when the cooling time is shorter than the halo dynamical time.  
	\item In our fiducial simulation ($Z=0.1\Zsun, b=0.3~\rm{kpc}$, prograde--prograde merger), 
	the mass of cold clouds ranges from $30~\Msun$ to $\sim 2.6 \times 10^{4}~\Msun$. 
	The upper bound is roughly regulated by the acoustic length in the post-shock region. 
	Some clouds merge into more massive ones with $\gtrsim 10^{4}\Msun$ within $2~\rm{Myr}$, exceeding the Jeans mass. Therefore these massive clouds are likely to collapse further, resulting in massive star clusters in high-$z$ dwarf galaxies. 
	\item Hydrogen molecules significantly suppress the formation of cold clouds via thermal instability. If $\HM$ abundance is higher than $f_{\rm H2} \sim 10^{-3}$, no cold clouds form in our simulation.  This is because the efficient  $\HM$ cooling brings the gas temperature down to $T<10^4$\,K, and the condition for thermal instability is no longer  satisfied as the shape of the cooling function flattens. We therefore suggest a critical H$_2$ abundance in Eq.~(\ref{eq:Hcrit}), below which the thermal instability can occur.   We also discuss the impact of inhomogeneous UV background field in Section~\ref{sec:h2depend}.
	\item  The formation efficiency of cold clouds sensitively depends on the orbital parameters and the inclination angles of colliding galaxies.  In the prograde--retrograde merger, an oblique shock forms and a shear flow is induced, which suppresses the formation of cold clouds.  This does not happen in the prograde--prograde merger. 
	\end{itemize}

Our results about the dependencies on metallicity and $\HM$ abundance indicate 
that the star formation history in dwarf galaxies can significantly change
depending on their formation sites. 
These environmental effects have not been investigated well 
even in current cosmological simulations (e.g., \citealt{Ricotti16, Kim17}),
due to the limitations in numerical resolution and sample sizes. 
In this work, by using isolated merging galaxies, we successfully followed the formation of cold clouds via thermal instability. 
Moreover, our simulations have shown the criteria of $t_{\rm cool} < t_{\rm dyn}$ for the cloud formation, which was out of the range of previous works using idealized colliding flows (e.g., \citealt{KI02, IO15}).

As shown in our results, the radiative cooling by [C{\sc ii}] line or $\HM$ controls the formation of cold clouds.
In the present work, we introduced the cooling functions of [C{\sc ii}] and $\HM$ assuming an ionization equilibrium.
However, the abundance of free electrons can be much higher, if the post-shock region is collisionally ionized due to higher shock velocity of $> 10~\rm km~s^{-1}$.  Such a situation might arise in the collisions of more massive galaxies. 
This can enhance the cooling rate at $T<10^{4}~\rm{K}$ significantly, because the residual free electrons excite the $\rm{[C_{I\hspace{-0.1em}I}]}$ emission \citep{KI00,Richings2014} and induce the formation of H$_2$ \citep{Shapiro87}.
Therefore we need to solve the non-equilibrium chemical reactions and hydrodynamics consistently for a more accurate estimate of cooling rates. 
In addition, the higher cooling rate will reduce the scale of thermal instability, which will require even more expensive numerical simulations with higher spatial resolution of $\ll 0.1~\rm pc$.
In our future work, we will present more comprehensive studies of cloud formation in galaxies with a wider mass range, and investigate the impact of star formation on the ISM in high-$z$ galaxies. 

%
%
%----------------------------------------------------------------------
%
% Acknowledge
%
%----------------------------------------------------------------------
\section*{Acknowledgments}
We are grateful to Drs. Kazuyuki Omukai, Kazunari Iwasaki, and Kengo Tomida for helpful comments.
The numerical simulations were  carried out on the computer cluster, {\tt{Draco}}, at the Frontier Research Institute for Interdisciplinary Sciences of Tohoku University and XC30 system at the Center for Computational Astrophysics (CfCA) of the National Astronomical Observatory of Japan.
KN acknowledges the support by JSPS KAKENHI  Grant Number JP17H01111.
%----------------------------------------------------------------------
%
% References
%
%----------------------------------------------------------------------
\bibliographystyle{mn}

%\bibliography{mn-jour,HY}

\begin{thebibliography}{36}
\expandafter\ifx\csname natexlab\endcsname\relax\def\natexlab#1{#1}\fi

%A
\bibitem[Agarwal et al.(2014)]{Agarwal14} Agarwal, B., Dalla Vecchia, C., Johnson, J.~L., Khochfar, S., \& Paardekooper, J.-P.\ 2014, \mnras, 443, 648 

\bibitem[Alonso-Herrero et al.(2000)]{Alonso-Herrero00} Alonso-Herrero, A., Rieke, G.~H., Rieke, M.~J., \& Scoville, N.~Z.\ 2000, \apj, 532, 845 

%B
%\bibitem[Balbus(1986)]{Balbus86} Balbus, S.~A.\ 1986, \apjl, 303, L79 

\bibitem[Balbus(1995)]{Balbus95} Balbus, S.~A.\ 1995, in Ferrara A., McKee C. F., Heiles C., Shapiro P. R.,
eds, ASP Conf. Ser. Vol. 80, The Physics of the Interstellar Medium and
Intergalactic Medium. Astron. Soc. Pac, San Francisco, p. 328

%\bibitem[Balbus \& Soker(1989)]{Balbus89} Balbus, S.~A., \& Soker, N.\ 1989, \apj, 341, 611 

%\bibitem[Bovill \& Ricotti(2009)]{Bovill09} Bovill, M.~S., \& Ricotti, M.\ 2009, \apj, 693, 1859 

%\bibitem[Bovill \& Ricotti(2011)]{Bovill11a} Bovill, M.~S., \& Ricotti, M.\ 2011, \apj, 741, 17

%\bibitem[Bovill \& Ricotti(2011)]{Bovill11b} Bovill, M.~S., \& Ricotti, M.\ 2011, \apj, 741, 18 

%\bibitem[Brodie \& Strader(2006)]{Brodie06} Brodie, J.~P., \& Strader, J.\ 2006, \araa, 44, 193 

\bibitem[Bate \& Burkert(1997)]{Bate97} Bate, M.~R., \& Burkert, A.\ 1997, \mnras, 288, 1060

%C
\bibitem[Chen et al.(2017)]{Chen17} Chen, P., Norman, M.~L, Xu, H., \& Wise, J.~H 2017, ArXiv e-prints

\bibitem[Corbelli et al.(1997)]{Corbelli97} Corbelli, E., Galli, D., \& Palla, F.\ 1997, \apjl, 487, L53 

\bibitem[Cox et al.(2006)]{Cox06} Cox, T.~J., Dutta, S.~N., Di Matteo, T., et al.\ 2006, \apj, 650, 791
%D
\bibitem[de Grijs et al.(2003)]{deGrijs03} de Grijs, R., Lee, J.~T., Clemencia Mora Herrera, M., Fritze-v.~Alvensleben, U., \& Anders, P.\ 2003, New Astron., 8, 155

\bibitem[Devecchi et al.(2012)]{Devecchi12} Devecchi, B., Volonteri, M., Rossi, E.~M., Colpi, M., \& Portegies Zwart, S.\ 2012, \mnras, 421, 1465 

\bibitem[Draine(2011)]{Draine11} Draine, B.~T.\ 2011, Physics of the Interstellar and Intergalactic Medium (Princeton, NJ: Princeton Univ. Press)


%E
%\bibitem[Elmegreen \& Efremov(1997)]{Elmegreen97} Elmegreen, B.~G., \& Efremov, Y.~N.\ 1997, \apj, 480, 235 
%F
%\bibitem[Fall \& Rees(1985)]{FR85} Fall, S.~M., \& Rees, M.~J.\ 1985, \apj, 298, 18 

\bibitem[Field(1965)]{Field65} Field, G.~B.\ 1965, \apj, 142, 531 

%\bibitem[Forbes \& Bridges(2010)]{Forbes2010} Forbes, D.~A., \& Bridges, T.\ 2010, \mnras, 404, 1203 
%G
\bibitem[Galli \& Palla(1998)]{Galli98} Galli, D., \& Palla, F.\ 1998, \aap, 335, 403


\bibitem[Geen et al.(2017)]{Geen17} Geen, S., Soler, J.~D., \& Hennebelle, P.\ 2017, \mnras, 471, 4844 

\bibitem[Genel et al.(2009)]{Genel09} Genel, S., Genzel, R., Bouch{\'e}, N., Naab, T., \& Sternberg, A.\ 2009, \apj, 701, 2002 

%H
\bibitem[Haardt \& Madau(2012)]{HM12} Haardt, F., \& Madau, P.\ 2012, \apj, 746, 125 

\bibitem[Hasegawa \& Semelin(2013)]{Hasegawa13} Hasegawa, K., \& Semelin, B.\ 2013, \mnras, 428, 154 

\bibitem[Hennebelle \& Audit(2007)]{Hennebelle07} Hennebelle, P., \& Audit, E.\ 2007, \aap, 465, 431

\bibitem[Hennebelle \& Chabrier(2008)]{Hennebelle08} Hennebelle, P., \& Chabrier, G.\ 2008, \apj, 684, 395

%\bibitem[Hills(1980)]{Hills80} Hills, J.~G.\ 1980, \apj, 235, 986 

%\bibitem[Hollenbach \& McKee(1989)]{HM89} Hollenbach, D., \& McKee, C.~F.\ 1989, \apj, 342, 306 

\bibitem[Hopkins et al.(2014)]{Hopkins14} Hopkins, P.~F., Kere{\v s}, D., O{\~n}orbe, J., et al.\ 2014, \mnras, 445, 581 

\bibitem[Hopkins(2015)]{Hopkins15} Hopkins, P.~F.\ 2015, \mnras, 450, 53 

%I
\bibitem[Iliev et al.(2006)]{Iliev06} Iliev, I.~T., Mellema, G., Pen, U.-L., et al.\ 2006, \mnras, 369, 1625 

\bibitem[Inoue \& Omukai(2015)]{IO15} Inoue, T., \& Omukai, K.\ 2015, \apj, 805, 73 

%\bibitem[Inoue et al.(2016)]{Inoue16} Inoue, A.~K., Tamura, Y., Matsuo, H., et al.\ 2016, Science, 352, 1559 

%J
\bibitem[Jaacks et al.(2013)]{Jaacks13} Jaacks, J., Thompson, R., \& Nagamine, K.\ 2013, \apj, 766, 94 

\bibitem[Johnson et al.(2013)]{Johnson13} Johnson, J.~L., Dalla Vecchia, C., \& Khochfar, S.\ 2013, \mnras, 428, 1857 

\bibitem[Johnson et al.(2014)]{Johnson14} Johnson, J.~L., Whalen, D.~J., Agarwal, B., Paardekooper, J.-P., \& Khochfar, S.\ 2014, \mnras, 445, 686 


%K
%\bibitem[Kang et al.(1990)]{Kang90} Kang, H., Shapiro, P.~R., Fall, S.~M., \& Rees, M.~J.\ 1990, \apj, 363, 488 
\bibitem[Katz et al.(2015)]{Katz15} Katz, H., Sijacki, D., \& Haehnelt, M.~G.\ 2015, \mnras, 451, 2352 

\bibitem[Kim et al.(2017)]{Kim17} Kim, J.-h., Ma, X., Grudi{\'c}, M.~Y., et al.\ 2017, ArXiv e-prints

\bibitem[Kimm \& Cen(2014)]{Kimm14} Kimm, T., \& Cen, R.\ 2014, \apj, 788, 121 

%\bibitem[Kimm et al.(2015)]{Kimm15} Kimm, T., Cen, R., Devriendt, J., Dubois, Y., \& Slyz, A.\ 2015, \mnras, 451, 2900 

\bibitem[Koyama \& Inutsuka(2000)]{KI00} Koyama, H., \& Inutsuka, S.\ 2000, \apj, 532, 980 

\bibitem[Koyama \& Inutsuka(2002)]{KI02} Koyama, H., \& Inutsuka, S.\ 2002, \apjl, 564, L97

%\bibitem[Kruijssen(2015)]{Kruijssen15} Kruijssen, J.~M.~D.\ 2015, \mnras, 454, 1658 

%L
%M
%\bibitem[Maji et al.(2016)]{Maji16} Maji, M., Zhu, Q., Li, Y., et al.\ 2016, arXiv:1606.07091
\bibitem[Ma et al.(2015)]{Ma15} Ma, X., Kasen, D., Hopkins, P.~F., et al.\ 2015, \mnras, 453, 960  

\bibitem[Maio et al.(2011)]{Maio11} Maio, U., Khochfar, S., Johnson, J.~L., \& Ciardi, B.\ 2011, \mnras, 414, 1145 

\bibitem[McKee \& Ostriker(1977)]{MO77} McKee, C.~F., \& Ostriker, J.~P.\ 1977, \apj, 218, 148

%\bibitem[Micic et al.(2013)]{Micic13} Micic, M., Glover, S.~C.~O., Banerjee, R., \& Klessen, R.~S.\ 2013, \mnras, 432, 626 

%N
\bibitem[Navarro et al.(1995)]{Navarro95} Navarro, J.~F., Frenk, C.~S., \& White, S.~D.~M.\ 1995, \mnras, 275, 56 

%O
\bibitem[Omukai(2000)]{Omukai00} Omukai, K.\ 2000, \apj, 534, 809 

\bibitem[Okamoto et al.(2008)]{Okamoto08} Okamoto, T., Gao, L., \& Theuns, T.\ 2008, \mnras, 390, 920 

%P
\bibitem[Perret et al.(2014)]{Perret14} Perret, V., Renaud, F., Epinat, B., et al.\ 2014, \aap, 562, A1 

%Q
%R
%\bibitem[Renaud et al.(2014)]{Renaud14} Renaud, F., Bournaud, F., Kraljic, K., \& Duc, P.-A.\ 2014, \mnras, 442, L33 

%\bibitem[Renaud et al.(2015)]{Renaud15} Renaud, F., Bournaud, F., \& Duc, P.-A.\ 2015, \mnras, 446, 2038 

\bibitem[Richings et al.(2014)]{Richings2014} Richings, A.~J., Schaye, J., \& Oppenheimer, B.~D.\ 2014, \mnras, 442, 2780 

\bibitem[Ricotti et al.(2016)]{Ricotti16} Ricotti, M., Parry, O.~H., \& Gnedin, N.~Y.\ 2016, \apj, 831, 204 

%S
\bibitem[Shapiro \& Kang(1987)]{Shapiro87} Shapiro, P.~R., \& Kang, H.\ 1987, \apj, 318, 32 

\bibitem[Smith et al.(2016)]{Smith16} Smith, A., Bromm, V., \& Loeb, A.\ 2016, \mnras, 460, 3143 

\bibitem[Springel et al.(2005)]{Springel05} Springel, V., Di Matteo, T., \& Hernquist, L.\ 2005, \mnras, 361, 776 

\bibitem[Susa(2008)]{Susa08} Susa, H.\ 2008, \apj, 684, 226

\bibitem[Susa \& Umemura(2004)]{Susa04} Susa, H., \& Umemura, M.\ 2004, \apj, 600, 1 


%T
\bibitem[Takahashi et al.(1983)]{Takahashi83} Takahashi, T., Silk, J., \& Hollenbach, D.~J.\ 1983, \apj, 275, 145 

%\bibitem[Tegmark et al.(1997)]{Tegmark97} Tegmark, M., Silk, J., Rees, M.~J., et al.\ 1997, \apj, 474, 1 
%\bibitem[Trenti et al.(2009)]{Trenti09} Trenti, M., Stiavelli, M., \& Michael Shull, J.\ 2009, \apj, 700, 1672 

\bibitem[Trenti et al.(2015)]{Trenti15} Trenti, M., Padoan, P., \& Jimenez, R.\ 2015, \apjl, 808, L35 
%U
%V
\bibitem[V{\'a}zquez-Semadeni et al.(2007)]{Vazquez-Semadeni2007} V{\'a}zquez-Semadeni, E., G{\'o}mez, G.~C., Jappsen, A.~K., et al.\ 2007, \apj, 657, 870 

%X
%Y
\bibitem[Yajima et al.(2012)]{Yajima12} Yajima, H., Choi, J.-H., \& Nagamine, K.\ 2012, \mnras, 427, 2889 

\bibitem[Yajima et al.(2011)]{Yajima11} Yajima, H., Choi, J.-H., \& Nagamine, K.\ 2011, \mnras, 412, 411 

\bibitem[Yajima et al.(2014)]{Yajima14} Yajima, H., Li, Y., Zhu, Q., et al.\ 2014, \mnras, 440, 776 

\bibitem[Yajima \& Khochfar(2016)]{Yajima16} Yajima, H., \& Khochfar, S.\ 2016, \mnras, 457, 2423

\bibitem[Yajima et al.(2017)]{Yajima17} Yajima, H., Nagamine, K., Zhu, Q., Khochfar, S., \& Dalla Vecchia, C.\ 2017, \apj, 846, 30 

%W
%\bibitem[Watson et al.(2015)]{Watoson15} Watson, D., Christensen, L., Knudsen, K.~K., et al.\ 2015, \nat, 519, 327 
\bibitem[Wise et al.(2012)]{Wise12a} Wise, J.~H., Turk, M.~J., Norman, M.~L., \& Abel, T.\ 2012a, \apj, 745, 50 

\bibitem[Wise et al.(2012)]{Wise12b} Wise, J.~H., Abel, T., Turk, M.~J., Norman, M.~L., \& Smith, B.~D.\ 2012b, \mnras, 427, 311 

\bibitem[Wise et al.(2014)]{Wise14} Wise, J.~H., Demchenko, V.~G., Halicek, M.~T., et al.\ 2014, \mnras, 442, 2560 

%\bibitem[Wolfire et al.(1995)]{Wolfire95} Wolfire, M.~G., Hollenbach, D., McKee, C.~F., Tielens, A.~G.~G.~M., \& Bakes, E.~L.~O.\ 1995, \apj, 443, 152

%Z

\end{thebibliography}

%----------------------------------------------------------------------
%
% Appendix
%
%----------------------------------------------------------------------
\appendix

%\section{Resolution study}

\section{Colliding flow tests}
\label{sec:appA}

Our work is motivated by the simple colliding flow simulation performed by \cite{IO15}, who used a mesh-based hydrodynamic simulations and showed the formation of two-phase ISM structure.  
They argued that the acoustic length ($\un{\ell}{ac}$) should be resolved by more than $\sim 64$ grids 
in order to follow the formation of cold clouds via thermal instability. 
Here we check whether our current resolution is sufficient to study the two-phase structure. 

The resolution in the Lagrangian scheme is determined by the particle mass ($\un{m}{gas}$) and the number of neighbor particles ($\un{N}{ngb}$).
Under the fixed number of neighbor particles $\un{N}{ngb}=32$, 
we compare three simulations with different mass resolution of $\un{m}{gas}=0.3\,\Msun$ (our fiducial simulation:~Sec.~\ref{sec:result}), $3.0\,\Msun$,~and~$30\,\Msun$
using a similar initial condition as \cite{IO15}. 
The initial density is $\un{n}{H}=2.5\ \cc$, the relative velocity is $|V|=16\ \rm{km~s^{-1}}$, and the metallicity is  $Z=0.1~\Zsun$.
We then add 10 per cent random perturbations to the density field. 

\begin{figure}
\begin{center}
\includegraphics[width=\columnwidth, bb=0 0 563 963]{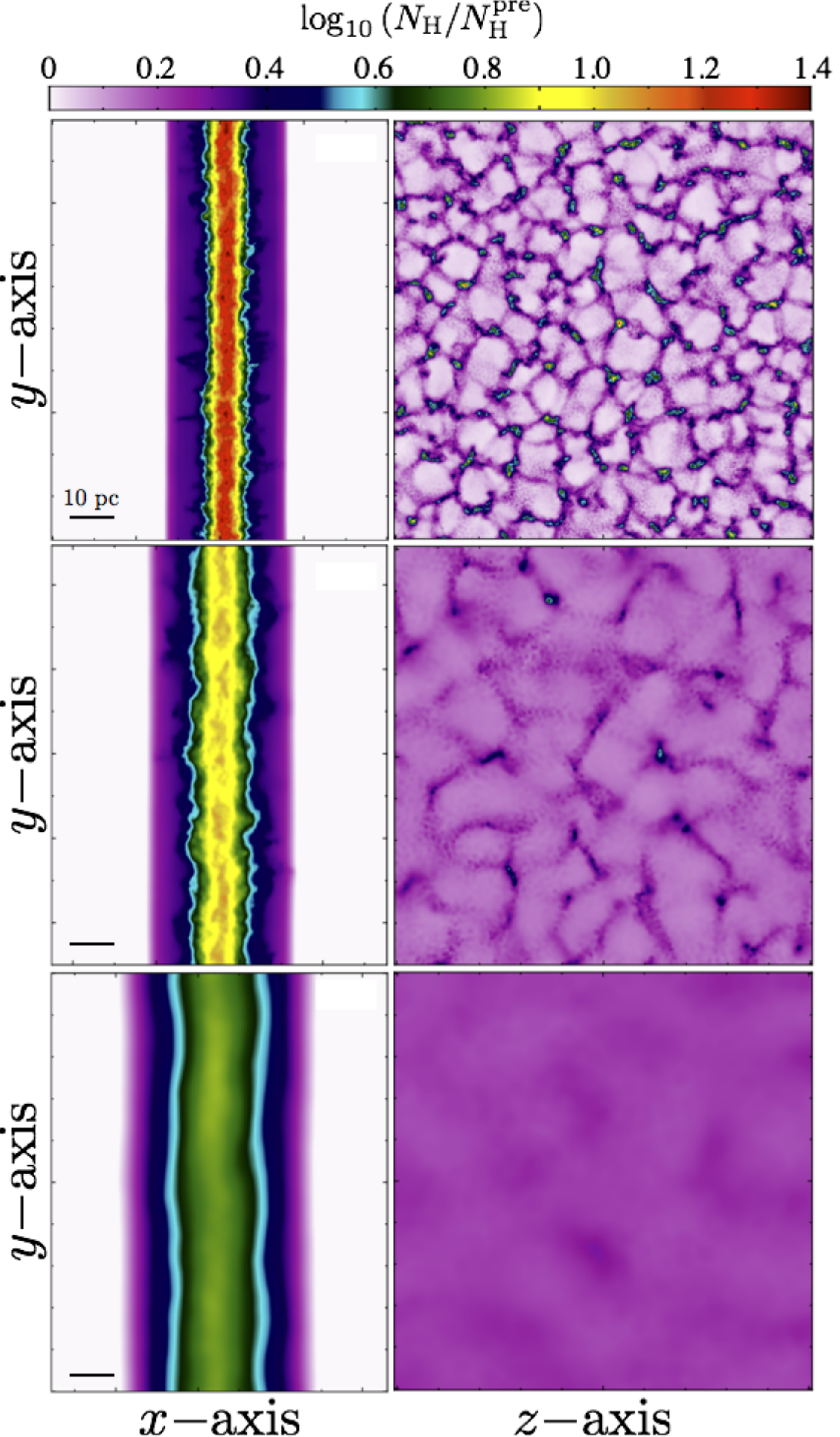}
\caption{
Column density maps (x-y and y-z plane) of the colliding flow tests. 
From top to bottom panels, the mass resolution is $\un{m}{gas}=0.3\,\Msun,~3.0\,\Msun,$ and $30\,\Msun$, respectively. 
The color bar at the top indicates the range of gas column density, and the normalization factors are $N_{\rm H}^{\rm pre} = 6.7\times 10^{21}, 1.4\times 10^{22}, 3.1\times 10^{22}$\,[cm$^{-2}$] for top to bottom panels, respectively.  
The black bar in the bottom-left corner of the left column panels represents $10~{\rm pc}$ (physical). 
The velocity of gas flow is initially given along the x-axis.
These snapshots show the density distribution at $t\approx 3.0~\un{t}{cool}$.
The cold clouds form via thermal instability within the acoustic length ($\un{\ell}{ac}\approx 16\ \rm{pc}$)
in the top two panels. 
}
\label{fig:coldflow}
\end{center}
\end{figure}

Figure~\ref{fig:coldflow} presents the column density maps at $t = 3.0 ~\un{t}{cool}$. 
The area shown in red or yellow color represent the post-shock region.
The top two panels show inhomogeneous and clumpy structure,
while in the bottom panel the post-shock region is almost homogeneous.

\begin{figure}
\begin{center}
\includegraphics[width=\columnwidth, bb=0 0 739 534]{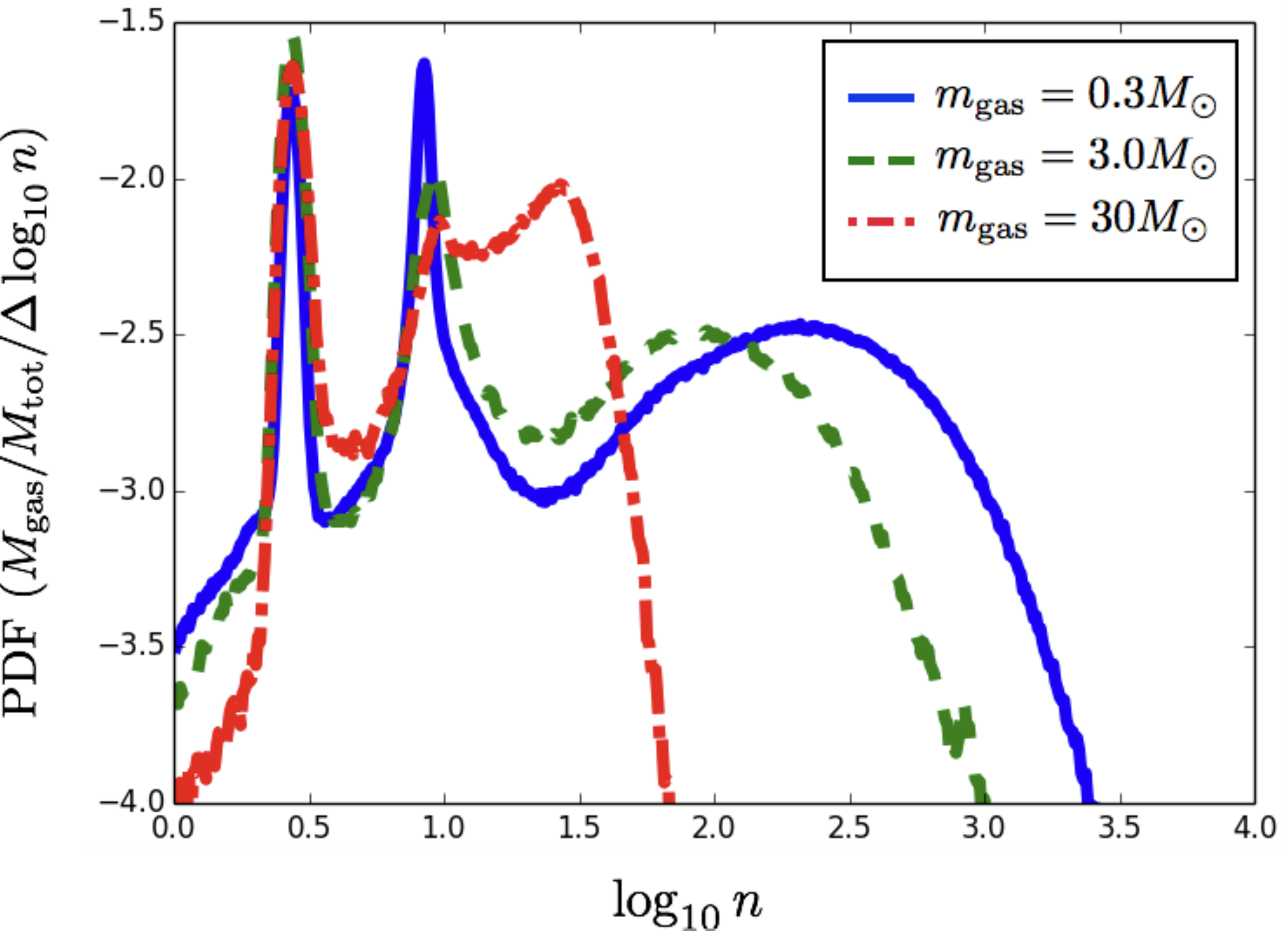}
\caption{
Probability distribution function of hydrogen number density [$\cc$] at $t\approx 3.0~\un{t}{cool}$ in the colliding flow tests.
Their mass resolution is $\un{m}{gas}=0.3\,\Msun$\,(red dot-dashed), $3.0\,\Msun$\,(green), and $30\,\Msun$\,(blue).
The first peak ($n \sim 2.5\ \cc$) represents the pre-shock gas. 
The post-shock gas shows two peaks at $n> 8.0\ \cc$ in the two higher-resolution cases owing to thermal instability, while they become weaker in the lowest resolution case (red dot-dashed).
}
\label{fig:resolution}
\end{center}
\end{figure}

Figure~\ref{fig:resolution} shows the probability distribution function (PDF) of gas density for three simulations. 
The PDF for the case with $\un{m}{gas}=0.3\,\Msun$ shows three peaks. 
The first peak ($\log_{10}{n [{\rm \cc}]}\approx 0.5$) represents the pre-shock gas, i.e., the initial background gas. 
The second peak ($\log_{10}{n} \approx 1$) shows the post-shock region which is compressed by a factor of $\sim 4$.
The third peak ($\log_{10}{n}\approx2.5$) indicates the cold clouds induced by thermal instability.
These features in the PDF are close to those in \citet{IO15}. 

The PDF for the case with $\un{m}{gas}=3.0 ~\Msun$ also shows
the three peaks. Even with this intermediate mass resolution, 
the acoustic length is resolved, resulting in the formation of massive cold clumps. 
However, the gas mass at the highest density $\log_{10}{n}\gtrsim 2 $ is
much lower than the case with $\un{m}{gas}=0.3\,\Msun$.
This is because the perturbations with shorter wavelengths 
are suppressed due to the weaker pressure gradient force with a coarser resolution. 

On the other hand, the lowest resolution run with $\un{m}{gas}=30\,\Msun$
does not show the highest density peak clearly, because the acoustic length is not resolved well. 
Therefore, we find that the mass resolution of gas particles should be lower than $\sim 10^{-3}$ of 
the maximum cloud mass ($M_{\rm max}$, Eq.\,\ref{eq:estimate}) formed by the growth of perturbation of the acoustic length in order to resolve the two-phase medium. 
% \sim \frac{4\pi}{3} \rho (\frac{\un{\ell}{ac}}{2})^{3}$, 

%%Fig. A1

%%Fig.A2

\label{lastpage}

\end{document}